\documentclass[letterpaper,12pt]{JHEP3}

\usepackage{amssymb,latexsym,bbm,kec,epsfig}

\title{Proteins Wriggle}

\author{Michael Cahill \\
	School of Medicine, Uniformed Services University\\
	Bethesda, Maryland 20814\\
	E-mail: \email{s4mcahill@usuhs.mil}}

\author{Sean Cahill \\
	Department of Computer Science,
	University of New Mexico\\
	Albuquerque, NM 87131\\
	E-mail: \email{locky@cs.unm.edu}}

\author{Kevin Cahill \\ 
        Department of Physics and Astronomy,
	University of New Mexico\\
	Albuquerque, New Mexico 87131-1156\\
	E-mail: \email{cahill@unm.edu}}
\preprint{cond-mat/0108218}

\abstract{We propose an algorithmic strategy 
for improving the efficiency of Monte Carlo searches
for the low-energy states of proteins.
Our strategy is motivated by a model
of how proteins alter their shapes.
In our model 
when proteins fold under physiological conditions,
their backbone dihedral angles 
change synchronously in groups of 
four or more 
so as to avoid steric clashes and respect the
kinematic conservation laws.
They wriggle; they do not thrash.
We describe a simple algorithm that 
can be used to incorporate wriggling
in Monte Carlo simulations of protein folding.
We have tested this wriggling algorithm 
against a code in which the dihedral angles
are varied independently (thrashing). 
Our standard of success is the average root-mean-square
distance (rmsd) between the \(\alpha\)-carbons
of the folding protein and those of its native structure.
After 100,000 Monte Carlo sweeps,
the relative decrease in the mean rmsd, 
as one switches from thrashing to wriggling,
rises from 11\%
for the protein 3LZM with 164 amino acids (aa) 
to 40\% for the protein 1A1S with 313 aa
and 47\% for the protein 16PK with 415 aa.
These results suggest that wriggling
is useful and that its
utility increases with the size of the protein.
One may implement
wriggling on a parallel computer or
a computer farm.}

\keywords{protein folding, Monte Carlo searches}

\begin{document}

\section{Why Proteins Wriggle}
We propose an algorithmic strategy 
for improving the efficiency of Monte Carlo searches
for the low-energy states of proteins.
Our strategy is motivated by a model in which
proteins alter their shapes by means of
local motions that minimize the displacement
of distant atoms.
\par
Folding proteins avoid steric clashes and 
respect the kinematic conservation laws.
The system consisting
of a protein and the nearby solvent molecules
approximately conserves its
energy, momentum, and angular momentum.
The shape of a protein is mainly 
defined by the angles
of rotation, $\phi_i$ and $\psi_i$, 
about the backbone bonds
that link the \(\alpha\)-carbons to the
adjacent amide planes.
A change in a one of these dihedral angles
would rotate a significant part of 
the protein molecule,
moving each atom by a length 
proportional to its distance from the axis of rotation.
In a protein consisting of
hundreds or thousands of amino acids
in water, such a rotation would engender
steric clashes and grossly violate the kinematic
conservation laws.
Instead when a protein folds or unfolds in our model,
its backbone dihedral angles 
conspire in groups of four or more
to change in ways that
limit their displacement of distant atoms.
Proteins wriggle; they do not thrash.
\par
Localized motions of the protein backbone
involve at least four bonds,
but simpler local motions are possible 
in simpler systems.
In lattice models all local motions,
such as corner moves and crankshaft moves~\cite{Schatzki1965},
are localized.
Polymers also possess simple
local motions in the continuum.
In polyethylene, for instance, 
two backbone bonds separated
by a \emph{trans} bond are parallel, 
and so equal and opposite
rotations about these parallel bonds
constitute a motion that is localized. 
Such crankshaft moves in polymers have 
been seen in simulations guided by brownian and molecular
dynamics~\cite{Helfand1971,Skolnick&Helfand1980,
Helfand&Wasserman&Weber1980,
Helfand&Wasserman&Weber1981,
Helfand&Wasserman&Weber&Skolnick&Runnels1981,
Weber&Helfand&Wasserman1983,Helfand1984}.
\par
Continuum Monte Carlo searches strike a balance 
between the temporal detail of molecular dynamics
and the rigidity of the lattice.
They are defined by their 
kinematics (how the proteins move)
and their dynamics (why they move).
By \emph{kinematics} we mean 
the variables that define the state of the protein
and the kinds of Monte Carlo moves that are permitted.
The \emph{dynamics} is determined by the energy function.
Monte Carlo simulations do not incorporate wriggling
in their kinematics;
this paper is about how 
they could and whether they should.
\par
Our wriggling algorithm is based upon the linear dependence
of every quartet of three-dimensional vectors.
In the next section, we use this linear dependence to show
that one may choose the four angles of rotation
about any four backbone bonds so that the combined motion
of the protein is localized.
A simple computer algorithm that may be used
to incorporate wriggling in Monte Carlo searches
is outlined in section 3.
\par
We have run three simple tests to determine 
whether wriggling actually
improves the efficiency of a Monte Carlo search;
we describe these tests and their results in section 4.
In each test we compared simulations guided 
by the wriggling algorithm to ones guided 
by a standard thrashing algorithm 
in which the dihedral angles are varied independently. 
In order to separate the kinematics of folding
from the dynamics of folding,
we used a nearly perfect but artificial energy function
that is proportional to the root-mean-square distance 
(rmsd) of the folding \(\alpha\)-carbons
from the \(\alpha\)-carbons of the native structure. 
Because there is no simple relation between this rmsd
and an energy, and because the use of wriggling 
alters the effective temperature,
we performed our Monte Carlo runs at absolute zero.
On each protein
we performed two sets of eight or more runs of 100,000 sweeps,
one set controlled by the wriggling algorithm and the other
by a thrashing algorithm.  The runs started
from fully denatured random coils. 
We averaged the final rmsd's.
The relative decrease in the average final rmsd
as one switches from the thrashing code to 
the wriggling code
\( \left( \langle \mathrm{rmsd} \rangle_\mathrm{th} 
- \langle \mathrm{rmsd} \rangle_\mathrm{wr} \right)/
\langle \mathrm{rmsd} \rangle_\mathrm{wr}\)
is a measure of the utility of wriggling.
This wriggling advantage rose
from 11\% for the protein 3LZM with 164 amino acids (aa),
to 40\% for the protein 1A1S with 313 aa,
and to 47\% for the protein 16PK with 415 aa.
The advantage of wriggling seems to grow
with the length of the protein.
\par
In section 5 
we sketch how one might implement wriggling
with a realistic energy function
on a parallel computer or on a farm of computers.
We summarize the present work and
mention some of its limitations in section 6.

\section{How Proteins Wriggle}
Three-dimensional space is spanned by any three 
linearly independent vectors.
Every quartet of three-dimensional vectors
is linearly dependent --- that is,
for every four three-dimensional vectors
\(\vec v_1, \vec v_2, \vec v_3, \vec v_4\), 
there exist four numbers
\(x_1, x_2, x_3, x_4\) such that 
the weighted sum of the vectors vanishes,
\beq
\sum_{i=1}^4 x_i \, \vec v_i = 0 .
\label{lindep}
\eeq
In this section we use 
these mathematical facts to show that one always may
choose the angles of rotation about any four bonds
so as to minimize the net effect of the four rotations
upon distant atoms. 
\par
The change \(\vec dr\) in the position \(\vec r\)
of an atom due to a rotation 
by a small angle \(\epsilon\) about 
a bond axis taken to be a 
unit vector \(\hat b\) is 
the cross-product of \(\epsilon \hat b\)
with the vector from any point \(\vec c\)
on the axis to the point \(\vec r\), 
\beq
\vec dr = \epsilon \hat b \times ( \vec r - \vec c ).
\label{dr=}
\eeq
To first order in \(\epsilon\),
the net displacement \(\vec dr\) of 
the position \(\vec r\) of an atom 
due to four rotations by the small angles 
\(\epsilon_i\)
about the bonds 
\(\hat b_i\) for \(i = 1,2,3,4\) 
is the sum
\beq
\vec dr = \sum_{i=1}^4 \vec dr_i
= \sum_{i=1}^4 
\left( \epsilon_i \hat b_i \times ( \vec r - \vec c_i ) \right).
\label{sumdr1}
\eeq
If we use \(\vec a\) for the average
of the four points \(\vec c_i\), 
which typically would be the midpoint
between two \(\alpha\)-carbons, 
then we may express the net displacement 
\(\vec dr\) as
\beq
\vec dr = \sum_{i=1}^4 
\left( \epsilon_i \hat b_i \times 
( \vec r - \vec a + \vec a - \vec c_i ) \right) 
= \left( \sum_{i=1}^4 
\epsilon_i \hat b_i \right) \times (\vec  r - \vec a )
+  \sum_{i=1}^4 
\left( \epsilon_i \hat b_i \times 
( \vec a - \vec c_i ) \right)  .
\label{sumdr2}
\eeq
The displacement \(\vec dr\)
will be independent of the potentially large
moment arm \(\vec r - \vec a\) (for every atom) 
if the sum of the bond vectors \(\hat b_i\) 
weighted by their angles \(\epsilon_i\) vanishes.  
That is, the net displacement \(\vec dr\) 
is merely 
\beq
\vec dr = \sum_{i=1}^4 
\left( \epsilon_i \hat b_i \times 
( \vec a - \vec c_i ) \right),  
\label{sumdr3}
\eeq
which is independent of \(\vec r - \vec a\),
if the angle-weighted sum of bond vectors vanishes,
\beq
\sum_{i=1}^4 
\epsilon_i \hat b_i = 0.
\label{sum=0}
\eeq
And because every quartet of three-dimensional 
vectors is linearly dependent (\ref{lindep}), 
it is always possible to choose the four small angles
\(\epsilon_i\) so that this sum vanishes
for any four bond vectors \(\hat b_i\)\@.

\section{A Wriggling Algorithm}
In this section we show how to transform
the wriggling condition (\ref{sum=0}) 
into a matrix equation that can be solved
by standard linear-algebra software,
such as the freely available LAPACK~\cite{LAPACK}
subroutine DGESV~\cite{DGESV}\@. 
\par
The wriggling condition (\ref{sum=0})
can be written in the more explicit form
\beq
\sum_{n=1}^3 \, \hat b_{in} \, (- \epsilon_n / \epsilon_4 ) = \hat b_{i4} 
\label{sum3=4}
\eeq
for \( i = 1, 2, 3\)\@.
Let us arrange the first three
axes $\hat b_1$, $\hat b_2$, and $\hat b_3$ 
into the matrix \(A\) 
with elements \( A_{in} = \hat b_{in} \) 
for \(i,n = 1, 2, 3\)
and rename the fourth axis $\hat b_4$ 
as the vector $B$ with components
\( B_i = \hat b_{i4} \)\@.
If we now use \(X\) for the vector with components
\(X_n = - \epsilon_n / \epsilon_4 \) for \(n = 1, 2, 3\),
then the wriggling condition (\ref{sum3=4}) becomes
\beq
\sum_{n=1}^3 \, A_{in} \, X_n = B_i 
\label {lineq}
\eeq
for $i = 1, 2, 3$\@.
\par
The LAPACK subroutine DGESV is designed to solve such 
linear equations. 
The call 
\beq
%\begin{verbatim} 
\mathtt{ 
call\ DGESV\ (\ 3,\ 1,\ A,\ 3,\ ipiv,\ B,\ 3,\ info\ )}
%\end{verbatim}
\eeq
returns the three angle ratios 
$ X_n = - \epsilon_n / \epsilon_4 $ for \(n = 1, 2, 3\)
as the three components \(B_n\)
of the vector \texttt{B}\@. 
The value \texttt{info = 0} indicates that
the computation is successful, and
\texttt{ipiv} contains pivot indices, 
which may be ignored.
We set \(\epsilon_4 = -1\) so that
\(\epsilon_n = - \epsilon_4 \, B_n = B_n\)
for \(n = 1, 2, 3\) and then
normalize the four angles
\beq
\sum_{n=1}^4 \epsilon_n^2 = 1.
\label{normeps}
\eeq
Lastly we multiply them
by a random number \(x\) drawn uniformly from
the interval \(( - 0.0125, 0.0125 )\) radians,
so that our final angles are \(\theta_n = x \epsilon_n\)
for \(n = 1, 2, 3, 4\)\@.
\par
Although we used the linear equation (\ref{sum3=4})
as our wriggling condition,
we used the exact and general form (\ref{Rcfc})
of the rotation matrix described
in the appendix to implement all rotations.

\section{Does Wriggling Work?}
In order to test the utility
of our wriggling algorithm,
we performed Monte Carlo 
simulations of protein folding
on three proteins: T4 lysozyme (3LZM.pdb, 164 aa),
ornithine carbamoyltransferase (1A1S.pdb, 313 aa),
and phosphoglycerate kinase (16PK.pdb, 415 aa)\@.
\par
We used an artificially nearly perfect 
energy function that is proportional 
to the root-mean-square distance (rmsd)
between the \(\alpha\)-carbons of the folding
protein and those of its native structure.
This nearly perfect energy function allows us to
separate the kinematics of folding
(the Monte Carlo moves --- wriggling or thrashing)
from the dynamics of folding
(the mechanisms in the energy function
--- conformational entropy, charge-charge interactions,
hydrogen bonds, van der Waals interactions,
hydrophobicity~\cite{Dill1990,Chan&Dill1990}).
Because there is no simple relationship
between the \(\alpha\)-carbon rmsd and an energy,
and because wriggling changes the effective
temperature,
we conducted our simulations
at absolute zero rather than at
physiological temperatures or
at that of liquid nitrogen.
\par
Each of our tests consisted of
8 or 10 pairs of Monte Carlo runs,
one guided by the wriggling algorithm
and the other by a thrashing algorithm. 
Each run began with a random coil 
and ran for 100,000 sweeps,
each sweep being a sequence of applications
of the algorithm successively along 
the primary structure of the protein.
The wriggling code applies the algorithm described in 
Eqs.(\ref{sum3=4}--\ref{normeps}) 
to successive quartets of dihedral angles.
After each wriggle the code performs
a Metropolis step;
since the temperature is zero,
the wriggle is accepted if and only if
it lowers the rmsd.
In each sweep the wriggling code wriggles firstly
the four dihedral angles 
\(\phi_2, \psi_2, \phi_3, \psi_3\) 
of residues 2 and 3, secondly 
the angles \(\psi_2, \phi_3, \psi_3, \phi_4\)
of residues 2, 3, and 4, thirdly
the angles \(\phi_3, \psi_3, \phi_4, \psi_4\) 
of residues 3 and 4, 
and continues in this way
down the chain to the penultimate residue. 
The code does not vary the \(\phi\) angle
of any proline residue. 
To keep these angles fixed,
the code performs one of several 
procedures when one or more proline residues
is involved in a wriggle.
In each sweep
the thrashing code successively 
and independently changes,
by a random angle \(\delta \theta\) drawn
uniformly from the interval
\((-0.0125, 0.0125)\) radians,
every dihedral
angle from the first \(\psi\) to the last \(\phi\),
except for the \(\phi\)'s of the prolines;
it accepts each change if and only if
the change lowers the rmsd. 
Apart from end effects,
the thrashing code makes
the same number of Monte Carlo judgments
per sweep as does the wriggling code.
\par
In our first test of wriggling,
we constructed 8 fully denatured random
coils of the protein 3LZM,
which has 164 residues. The rmsd's
of these denatured configurations ranged
from 18.2 to 128.5 \AA\@. 
In 8 runs of 100,000 sweeps,
the average rmsd was \(1.46 \pm 0.07\) \AA\ 
for the wriggling code and
\(1.62 \pm 0.05\) \AA\ for the thrashing code.
The mean thrashing rmsd was 11\% larger
than the mean wriggling rmsd. 
\begin{figure}
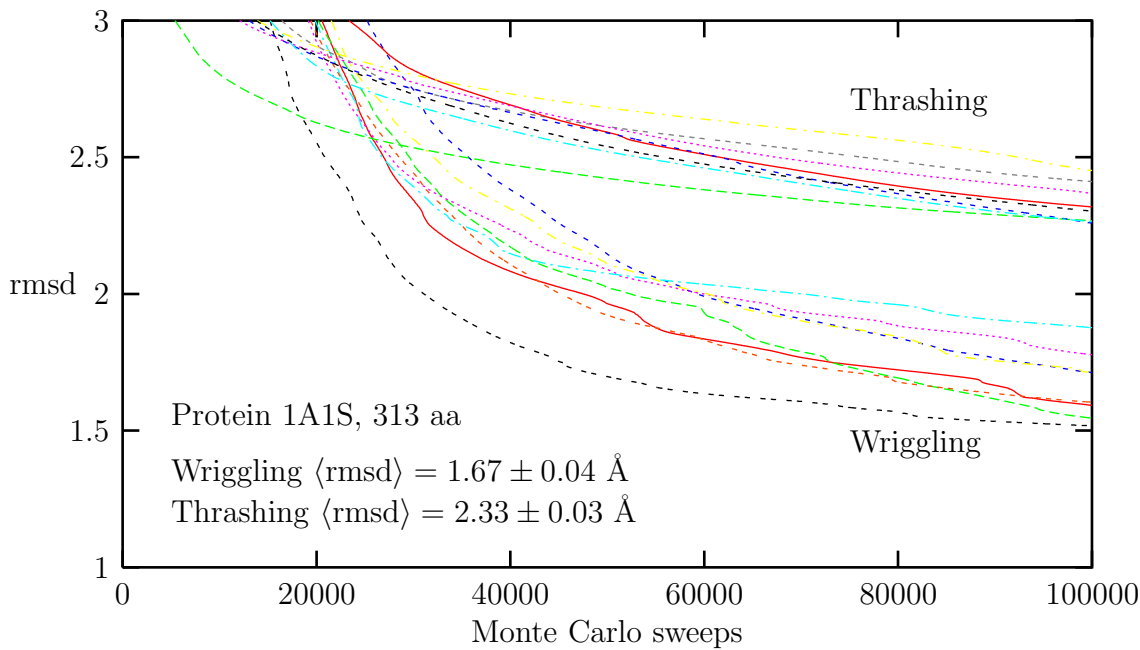

\centering
\input 1A1S
\caption{For the protein 1A1S,
the lines trace the values of the
rmsd for 8 runs guided by the wriggling
algorithm and 8 guided by the thrashing
algorithm.}
\label{1a1sfig}
\end{figure}
\par
We performed our second test 
of wriggling on the protein 1A1S,
which has 313 residues.
Our 8 denatured
coils of 1A1S had rmsd's
running from 25.9 \AA\ to 251.5 \AA\@.
The rmsd's of the 16 wriggling and thrashing runs 
are plotted in Fig.~\ref{1a1sfig}\@.
After about 40,000 sweeps, the thrashing
runs separate out into a cluster of lines,
labeled as thrashing,
that lie distinctly above the  
wriggling runs, labeled as wriggling.
The 8 wriggling runs had an average final rmsd of
\(1.67 \pm 0.04\) \AA, while that of the 8
thrashing runs was
\(2.33 \pm 0.03\) \AA\ or 40\% greater.
\begin{figure}
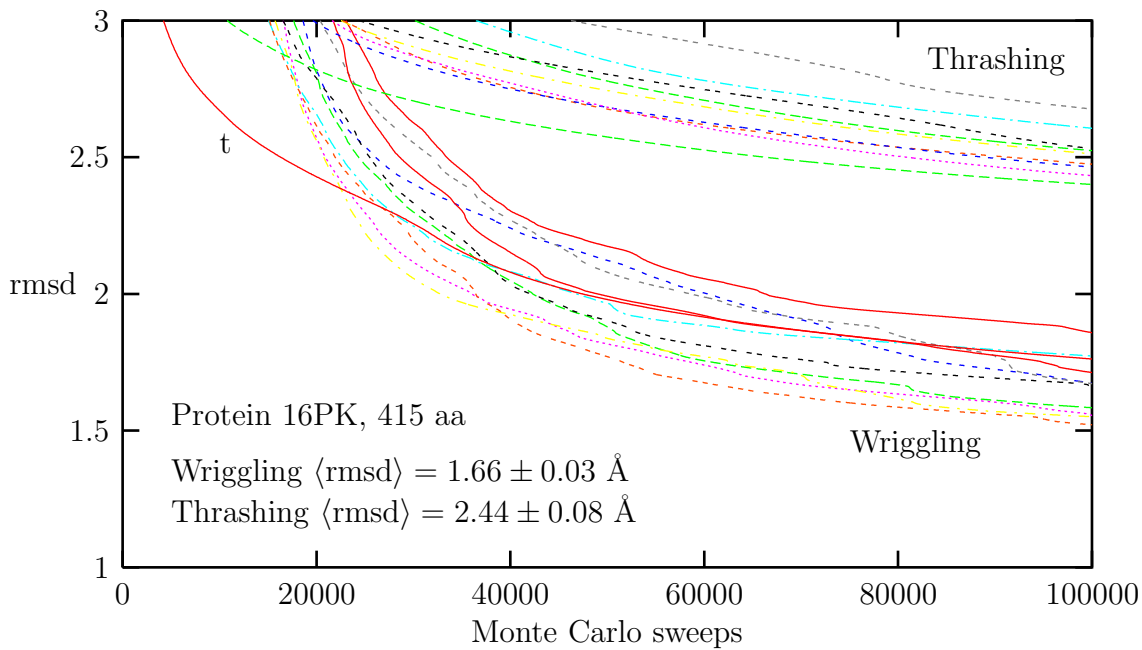

\centering
\input 16PK
\caption{For the protein 16PK,
the lines trace the values of the
rmsd for 10 runs guided by the wriggling algorithm 
and 10 guided by the thrashing algorithm. 
The plot labeled by the letter ``t'' 
is a successful thrashing outlier.}
\label{16pkfig}
\end{figure}
\par
For our third test of wriggling, 
we first randomized and stretched 
the native structure of the protein 16PK,
which has 415 (visible) residues,
into 10 fully denatured
coils with rmsd's
running from 24.7 \AA\ to 341.8 \AA\@.
We then allowed our wriggling and
thrashing codes to reduce the rmsd's
of these 10 random coils in runs
of 100,000 sweeps.
The rmsd's of the 20 runs
are plotted in Fig.~\ref{16pkfig} 
as a function of sweep number.
Apart from one thrashing run,
the rmsd's of
the wriggling runs drop below
those of the thrashing runs
after about 30,000 sweeps.
The outlying thrashing run,
labeled by the letter ``t,'' 
did slightly better than the two worst    
wriggling runs.
After 100,000 sweeps, the average rmsd
of the wriggling runs was 
\(1.66 \pm 0.03\) \AA,
while that of the thrashing runs was
\(2.44 \pm 0.08\) \AA\@.
The mean rmsd of the thrashing code
was 47\% greater than that of 
the  wriggling code.
\par
One estimate of the utility of wriggling
is the relative decrease in the average final rmsd
as one switches from the thrashing code to 
the wriggling code,
\beq
\frac{\langle \mathrm{rmsd} \rangle_\mathrm{th} 
- \langle \mathrm{rmsd} \rangle_\mathrm{wr}}
{\langle \mathrm{rmsd} \rangle_\mathrm{wr}}.
\label{utility}
\eeq
In these three tests,
the utility of wriggling 
increased with the size of the protein,
rising from an advantage 11\% at 164 aa
to 40\% at 313 aa and 
47\% at  415 aa. 
The longer the protein,
the larger are the motions that occur
when the backbone-bond angles are varied
one at a time, and so the greater are
the need for and the advantage of wriggling.
\par
Each of the proteins 1A1S and 16PK
has two domains.  Is the advantage of wriggling 
over thrashing in these two cases
due merely to a better twist in the
polypeptide strand that connects the two domains?
To answer this question, we measured the
rmsd's of the individual domains of 
the final configurations of 1A1S and 16PK
given by the wriggling and thrashing codes.
The average rmsd's of the first and second domains
of 1A1S respectively were 
1.78 and 1.56 \AA\ with wriggling
and 2.05 and 2.57 \AA\ with thrashing.
Those of 16PK were 
1.52 and 1.79 \AA\ with wriggling
and 2.43 and 2.67 \AA\ with thrashing.
These results suggest that
wriggling gives better domains, 
not just better connecting strands.
\par
The wriggling code differs from the 
thrashing code in two respects: 
its basic moves are four rotations 
rather than a single rotation and
large motions of remote atoms are suppressed.
To evaluate the two effects separately,
we wrote a code in which the elemental moves are
groups of four rotations but in which
no wriggling condition is imposed. 
We let this coordinated-thrashing code fold our
ten denatured starting configurations
of the protein 16PK and found 
after 100,000 sweeps that the 
average rmsd was \(1.88 \pm 0.04\)~\AA\,
which is to be compared with \(1.66 \pm 0.03\) \AA\
for the wriggling code and \(2.44 \pm 0.08\) \AA\ 
for the thrashing code.  So wriggling
is better than coordinated thrashing and
much better than thrashing,
but part of the success of wriggling
arises from the coordination of 
its compound elemental moves.
\par
Because of our use of the rmsd
as an artificially nearly perfect energy function,
the proteins of our simulations are phantoms;
they can move through each other.
A real but approximate energy function
would reject all moves into excluded volume;
it therefore would reject many thrashing moves 
because of their large-scale motions.
The use of the rmsd in our three tests  
deprives wriggling of one of its key advantages
over thrashing and over coordinated thrashing, 
namely that its localized
motions are less likely to involve steric clashes. 
Thus the utility of wriggling in simulations
with real energy functions may be greater 
than is indicated by these tests.
\par
The wriggling code runs somewhat more slowly
than the thrashing code.  
But a realistic energy function would slow
down both codes by so much that the
speed advantage of thrashing would be negligible. 
\par
In the first 10,000 sweeps 
of our tests of the wriggling algorithm,
the thrashing code reduced its rmsd's
more quickly than the wriggling code.
It might therefore be worthwhile
to experiment with codes that relax
the wriggling condition for the first 10,000 sweeps 
or that mix coordinated thrashing with wriggling.
\par
The action of a wriggle is much less
than that of a thrash;
the action of a small-angle wriggle 
may be as little as \(10\hbar\)\@. 
So quantum-mechanical effects
are more important with wriggling 
than with thrashing, but even so they probably 
would be obscured by decoherence~\cite{Zurek1998}.

\section{Wriggling on a Parallel Computer}
Most of the residues of
a folded protein lie in alpha helices 
and beta sheets.
The secondary structure of a protein
is the assignment of residues 
to helices, sheets, turns, and coils.  
One may list the possible secondary structures
of a protein and assign one secondary structure
to each processor of a parallel computer
or computer farm.
Each processor would perform
Monte Carlo moves on the dihedral angles
of the residues
in the turns and coils of its secondary structure
but would leave invariant the dihedral angles
of its helices and sheets.
Wriggling should be used in the coils
and in turns longer than 4 or 5 aa,
but probably not in turns of 3 or 4 aa,
where it might overly
constrain the folding of the protein.
Because each processor would vary
the dihedral angles (and possibly the 
principal side-chain angles) only of the residues
in the coils and turns,
the simulation would run quickly enough
to be guided by a realistic energy function
~\cite{Kollman1981,Karplus1983,Karplus1998,Karplus1999}
with solvation and excluded volume.
At the end of a run of perhaps 100,000 sweeps,
the final energies of the different 
secondary structures would be compared 
and their folds stored.
Many runs would be required to test
all the plausible secondary structures. 
This implementation of wriggling
would make optimum use of a parallel
computer or of a computer farm;
no time would be lost to inter-processor
communication or to waiting.

\section{Conclusions and Caveats}
We have described and tested an algorithmic strategy
for improving the efficiency of Monte Carlo searches
for the low-energy states of proteins.
Our strategy is motivated by a model in which
the laws of physics constrain 
the incremental motions
of proteins to be essentially local. 
Localized motions can be incorporated 
in Monte Carlo searches by a simple
algorithm that rotates the dihedral angles
in groups of four.  
To test this wriggling algorithm,
we performed 52 zero-temperature, 100,000-sweep, 
Monte Carlo searches for the low-energy states
of the proteins 3LZM, 1A1S, and 16PK
using the rmsd  
as an artificially nearly perfect energy function.
The searches guided by the wriggling algorithm
reached lower rmsd's than those guided by the 
usual thrashing algorithm by a margin
that increased with the length of the protein.
But it remains to be seen whether and how
this wriggling algorithm might improve the efficiency
of Monte Carlo searches performed
at finite temperature 
and guided by an approximate, 
realistic energy function
with solvation and excluded volume. 
 
\section{Acknowledgments}
We have benefitted from several conversations
with Susan Atlas, Ken Dill, Norman van Gulick,
and Gary Herling.
One of us (SC) would like
to thank Ken Dill for the hospitality
extended to him at UCSF.
Most of our computations were performed 
on the computers of the Albuquerque High-Performance
Computing Center.

\appendix

\section{Rotation Matrices}\label{sec:A}
For the sake of completeness,
we derive in this appendix an exact formula 
for the general rotation matrix 
from the expression (\ref{dr=}) for an
infinitesimal rotation.
\par
To simplify the notation,
we shall consider an axis that runs through
the origin and choose \(\vec c = 0\)\@.
In this case by Eq.(\ref{dr=}), 
a right-handed rotation about 
a bond \(\hat b\) by an infinitesimal angle \(\epsilon\)
changes a vector \(\vec r\) by the small amount
\(\vec dr = \epsilon \, \hat b \times \vec r\)
where the cross-product \(\hat b \times \vec r\) 
has the components 
\beq
(\hat b \times \vec r)_i = \sum_{j=1}^3 \, \sum_{k=1}^3 \,
\varepsilon_{ikj} \, \hat b_k \, r_j   
\label{crossproduct}
\eeq
in which the totally anti-symmetric tensor
\(\varepsilon_{ikj}\) has elements
\(\varepsilon_{123} = \varepsilon_{231} = \varepsilon_{312} = 1\),
\(\varepsilon_{213} = \varepsilon_{132} = \varepsilon_{321} = - 1\),
with all other elements zero,
\emph{e.g.} \(\varepsilon_{113} = 0, \emph{etc}\)\@.
If we use the definition 
\(
(L_k)_{ij} = \varepsilon_{ikj}
\label {Ldef}
\)
of the rotation generators \(\vec L\),
then we may write the change \(dr_i\) as 
\beq
dr_i =  \epsilon 
\sum _{j=1}^3 \sum _{k=1}^3  \hat b_k (L_k)_{ij} \, r_j 
= \epsilon \sum _{j=1}^3 
\left( \hat b \cdot \vec L \right)_{ij} \, r_j 
\label{dri}
\eeq
or in matrix notation as
\beq
dr = \epsilon \, \hat b \cdot \vec L \, r .
\label{dr}
\eeq
Let us use the \(\vec r(\theta)\) for the vector \(\vec r\)
after a right-handed rotation by the angle \(\theta\)
about the axis \(\hat b\)\@.
Then by Eq.(\ref{dr}) the vector \(\vec r(\theta)\)   
satisfies the differential equation
\beq
\frac{dr(\theta)}{d\theta} = 
\hat b \cdot \vec L \, r(\theta) .
\label{deq}
\eeq
The solution that satisfies the
boundary condition \( \vec r(0) = \vec r \) is
\beq
r(\theta) = \exp( \theta \hat b \cdot \vec L ) \, r(0),  
\label{rtheta}
\eeq
and so the matrix that represents a finite rotation
by the angle \(\theta\) about the axis \(\hat b\) is
\beq
R(\theta \hat b)  = \exp( \theta \hat b \cdot \vec L ). 
\label{R}
\eeq
\par
The exact form of Eq.(\ref{dr=}) 
when the axis \(\hat b\) does not 
go through the origin but through
another point \(\vec c\) is
\beq
r_i(\theta) - c_i = 
\sum_{j=1}^3 R_{ij}(\theta \hat b) \, ( r_j - c_j ) 
= \sum_{j=1}^3  
\left[ \exp( \theta \hat b \cdot \vec L)\right]_{ij} 
( r_j - c_j ).
\label{r'Rr}
\eeq
In our codes we used this formula
with the matrix \(R\) given by~\cite{Cahill2000}
\beq
R_{ij}(\theta \hat b) =
\cos \theta \, \delta_{ij} 
- \sin \theta \, 
\left( \sum_{k=1}^3 \epsilon_{ijk} \, \hat b_k \right)
+ ( 1 - \cos \theta ) \, \hat b_i \, \hat b_j
\label{Rcfc}
\eeq 
which is convenient for computation.

\bibliography{polymers,cs,proteins,coh}

\providecommand{\href}[2]{#2}\begingroup\raggedright\begin{thebibliography}{10}

\bibitem{Schatzki1965}
T.~F. Schatzki, {\it Molecular interpretation of the \(\gamma\)-transition in
  polyethylene and related compounds},  {\em Polymer Preprints (American
  Chemical Society, Division of Polymer Chemistry)} {\bf 6} (1965), no.~2
  646--651.

\bibitem{Helfand1971}
E.~Helfand, {\it Theory of the kinetics of conformational transitions in
  polymers},  {\em Journal of Chemical Physics} {\bf 54} (1971), no.~11
  4651--4661.

\bibitem{Skolnick&Helfand1980}
J.~Skolnick and E.~Helfand, {\it Theory of the kinetics of conformational
  transitions in polymers},  {\em Journal of Chemical Physics} {\bf 72} (1980),
  no.~10 5489--5500.

\bibitem{Helfand&Wasserman&Weber1980}
E.~Helfand, Z.~R. Wasserman, and T.~A. Weber, {\it Brownian-dynamics study of
  polymer conformational transitions},  {\em Macromolecules} {\bf 13} (1980)
  526--533.

\bibitem{Helfand&Wasserman&Weber1981}
E.~Helfand, Z.~R. Wasserman, and T.~A. Weber, {\it Kinetics of conformational
  transitions in bulk polymers},  {\em Polymer Preprints (American Chemical
  Society, Division of Polymer Chemistry)} {\bf 22} (1981) 279--280.

\bibitem{Helfand&Wasserman&Weber&Skolnick&Runnels1981}
E.~Helfand, Z.~R. Wasserman, T.~A. Weber, J.~Skolnick, and R.~J. H., {\it The
  kinetics of conformational transitions: Effect of variation of bond angle
  bending and bond stretching force constants},  {\em Journal of Chemical
  Physics} {\bf 75} (1981), no.~9 4441--4445.

\bibitem{Weber&Helfand&Wasserman1983}
T.~A. Weber, E.~Helfand, and Z.~R. Wasserman, {\em Simulation of Polyethylene},
  molecular-based study of fluids~20, pp.~487--500.
\newblock No.~204 in ACS Advances in Chemistry Series.
\newblock American Chemical Society, 1983.

\bibitem{Helfand1984}
E.~Helfand, {\it Dynamics of conformational transitions in polymers},  {\em
  Science} {\bf 226} (1984) 647--650.

\bibitem{LAPACK}
E.~Anderson, Z.~Bai, C.~Bischof, S.~Blackford, J.~Demmel, J.~Dongarra,
  J.~Du~Croz, A.~Greenbaum, S.~Hammarling, A.~McKenney, and D.~Sorensen, {\em
  LAPACK Users' Guide}.
\newblock SIAM, Philadelphia, PA, 3d~ed., 1999.
\newblock Available on line at
  http://www.netlib.org/lapack/lug/lapack\_lug.html.

\bibitem{DGESV}
E.~Anderson, Z.~Bai, C.~Bischof, S.~Blackford, J.~Demmel, J.~Dongarra,
  J.~Du~Croz, A.~Greenbaum, S.~Hammarling, A.~McKenney, and D.~Sorensen, {\em
  LAPACK Users' Guide}, pp.~237--238.
\newblock SIAM, Philadelphia, PA, 3d~ed., 1999.
\newblock Available on line at http://www.netlib.org/lapack/.

\bibitem{Dill1990}
K.~A. Dill, {\it Dominant forces in protein folding},  {\em Biochemistry} {\bf
  29} (1990) 7133--7155.

\bibitem{Chan&Dill1990}
H.~S. Chan and K.~A. Dill, {\it Origins of structure in globular proteins},
  {\em Proceedings of the National Academy of Sciences USA} {\bf 87} (1990)
  6388.

\bibitem{Zurek1998}
S.~Habib, K.~Shizume, and W.~H. Zurek, {\it Decoherence, chaos, and the
  correspondence principle},  {\em Physical Review Letters} {\bf 80} (1998)
  4361--4365, [\href{http://xxx.lanl.gov/abs/quant-ph/9803042}{{\tt
  quant-ph/9803042}}].

\bibitem{Kollman1981}
P.~K. Weiner and P.~A. Kollman, {\it Amber: Assisted model building with energy
  refinement. a general program for modeling molecules and their interactions},
   {\em Journal of Computational Chemistry} {\bf 2} (1981) 287.

\bibitem{Karplus1983}
B.~R. Brooks, R.~E. Bruccoleri, B.~D. Olafson, D.~J. States, S.~Swaminathan,
  and M.~Karplus, {\it Charmm: A program for macromolecular energy,
  minimization, and dynamics calculations},  {\em Journal of Computational
  Chemistry} {\bf 4} (1983), no.~2 187--217.

\bibitem{Karplus1998}
T.~Lazarides and M.~Karplus, {\it Discrimination of the native from misfolded
  protein models with an energy function including implicit solvation},  {\em
  Journal of Molecular Biology} {\bf 288} (1998) 477--487.

\bibitem{Karplus1999}
T.~Lazarides and M.~Karplus, {\it Effective energy function for proteins in
  solution},  {\em Proteins: Structure, Function, and Genetics} {\bf 35} (1999)
  133--152.

\bibitem{Cahill2000}
M.~Cahill, M.~Fleharty, and K.~Cahill, {\it Simulations of protein folding},
  {\em Nuclear Physics B (Proc.\ Suppl.)} {\bf 83} (2000) 929--931,
  [\href{http://xxx.lanl.gov/abs/hep-lat/9909080}{{\tt hep-lat/9909080}}].

\end{thebibliography}\endgroup
\bibliographystyle{JHEP}

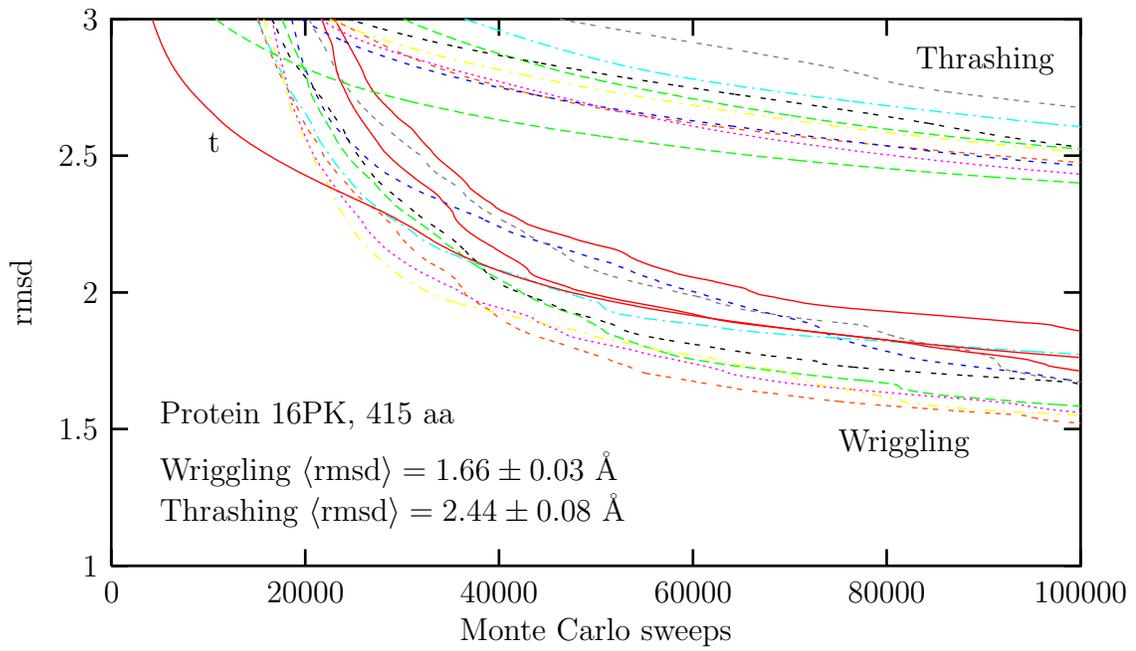
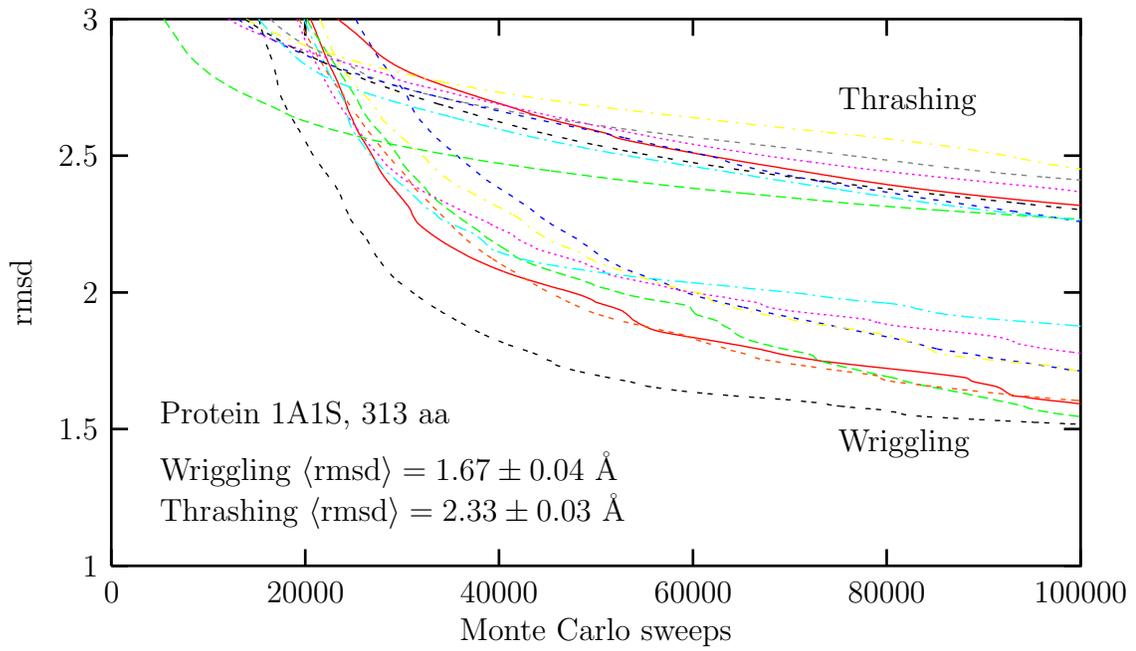
\end{document}